%% file: astroph.tex
\documentclass[apjl]{emulateapj}

\shorttitle{Discovery of Andromeda~XVII}
\shortauthors{}

\begin{document}

\title{Andromeda~XVII: A New Low Luminosity Satellite of M31}


\author{
  M.~J. Irwin\altaffilmark{1}, A.~M.~N. Ferguson\altaffilmark{2},  A.~P. Huxor\altaffilmark{2},  N.~R. Tanvir\altaffilmark{3}, R.~A. Ibata\altaffilmark{4}, G.~F. Lewis\altaffilmark{5}}
\altaffiltext{1}{Institute of Astronomy, University of Cambridge, Madingley Road, 
Cambridge, CB3 0HA, UK}
\altaffiltext{2}{Institute for Astronomy, University of Edinburgh, Royal Observatory, 
Blackford Hill, Edinburgh, EH9 3HJ, UK}
\altaffiltext{3}{Department of Physics \& Astronomy, University of Leicester, 
Leicester, LE1 7RH, UK}
\altaffiltext{4}{Observatoire de Strasbourg, 11 rue de l'Universit\'{e}, F-67000 Strasbourg, France}
\altaffiltext{5}{Institute of Astronomy, School of Physics, A29, University of Sydney, NSW 2006, Australia}

\begin{abstract} We report the discovery of a new dwarf spheroidal
  galaxy near M31 based on INT/WFC imaging. The system, Andromeda~XVII
  (And~XVII), is located at a projected radius of $\approx 44$~kpc
  from M31 and has a line-of-sight distance of 794$\pm$40 kpc measured
  using the tip of the red giant branch, and therefore lies well
  within the halo of M31.  The colour of the red giant branch implies
  a metallicity of [Fe/H]$\approx-1.9$ and we find an absolute
  magnitude of M$_V\approx-8.5$.  Three globular clusters lie near the
  main body of And~XVII, suggesting a possible association; if any of
  these are confirmed, it would make And~XVII exceptionally unusual
  amongst the faint dSph population. The projected position on the sky
  of And~XVII strengthens an intriguing alignment apparent in the
  satellite system of M31, although with a caveat about biases
  stemming from the current area surveyed to significant
  depth.  \end{abstract}


\section{Introduction}

The new generation of sensitive wide-field surveys has been rapidly
improving our knowledge of the satellite systems of the Milky Way and
M31.  These surveys are not only uncovering increasing levels of
complexity in the known dwarf spheroidal (dSph) population 
\citep[e.g.][]{komiyama07,mc07}, but are discovering many new low
surface brightness members of both systems.  In the case of the Milky
Way, fourteen new satellites have been reported in the last few years,
more than doubling the number of known and revealing a few systems
with properties more suggestive of extended star clusters than dwarf
galaxies
\citep[e.g.][]{willman05a,willman05b,belo06,saka06,zucker06a,zucker06b,belo07,koposov07,walsh07,irwin07}.
Progress has been similarily rapid for the M31 system, building on the
pioneering visual survey of \cite{vdb72}; the number of known dSph
systems has increased in recent years from six to fourteen, with the
nature of one system (And~VIII) still under dispute \citep{morrison03}
while a further candidate system, And~IV, has been conclusively shown
to be a background galaxy \citep{ferg00}.

Modern day M31 dSph discoveries have mainly come from a few panoramic
digital surveys.  This era began with \cite{armand98,armand99} and
\cite{kara99} using digitized POSS plates to discover And~V, VI and
VII. The SDSS survey led to the discovery of And~IX and X
\citep{zucker04,zucker07}.  A MegaCam CFHT survey uncovered And~XI,
XII, XIII, XV and XVI \citep{martin06,ibata07}, whilst And~XIV came
from a survey of the outermost part of the SE halo \citep{majewski07}.
With each new area surveyed more dwarfs are being uncovered and it
seems a complete census of both M31 and the Milky Way will contribute
to a resolution of the missing satellite problem
\citep[e.g.][]{geha07}.

We present here the discovery of a new dSph found in the course of
an extension of the INT/WFC imaging survey, previously described in
\cite{ferg02} and \cite{irwin05}.  Throughout this paper we assume a
distance to M31 of 785 kpc \citep{mc05}.

\section{Observations and Discovery}

For several years, we have been using the Isaac Newton 2.5~m Telescope
equipped with the Wide Field Camera (INT/WFC) to conduct an imaging
survey of M31 and the surrounding environment within a projected
radius of $\approx$60 kpc \citep{ibata01,ferg02,irwin05,huxor08}.
Observations are taken in the Johnson V and Gunn {\sl i} filters with
exposures in the range of 800-1200 seconds. Under typical seeing
conditions of $1.0-1.2 \arcsec$, this is sufficient to detect
individual stars at the distance of M31 to $\approx 3$ magnitudes
below the tip of the red giant branch (RGB).  The images are processed
using the Cambridge Astronomical Survey Unit (CASU) pipeline
\citep{irwin01} following the procedure outlined in \cite{ferg02} and
\cite{irwin05}. To date, some 200 contiguous fields have been
targetted, covering approximately 50 square degrees of sky centred on
M31, which at that distance corresponds to a projected surface area of
$\approx 100$~kpc $\times 100$~kpc.

The photometric calibration of the survey is based on a combination of
standard stars \citep{land92} observed on photometric nights and the
$\approx$10\% overlap between adjacent pointings.  Colour equations
defining the transformation between Landolt photometry and
instrumental V and {\sl i} magnitudes are given in
\cite{mc04}. Extinction-corrected values are based on the
\cite{schlegel98} extinction maps together with the extinction
coefficients defined therein.  Although the average value of the
extinction in this direction is quite low, E(B-V) $=$ 0.075, the
foreground extinction varies sufficiently to warrant star-by-star
correction as a function of field position and this has been applied
to the photometry reported in this paper.

\begin{figure}[h]
\begin{center}
\includegraphics[width=75mm]{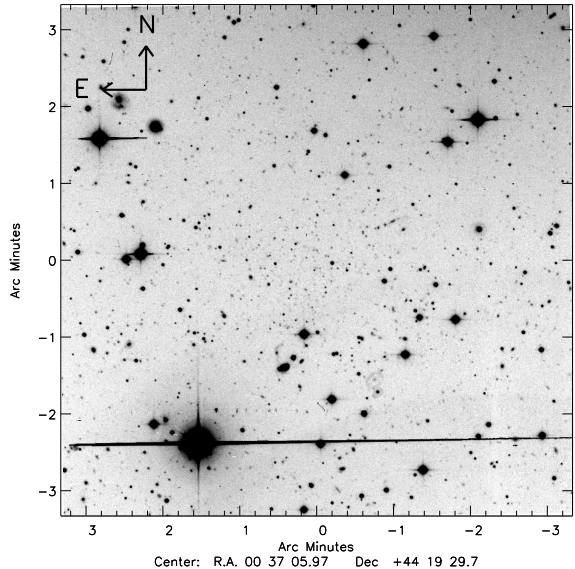}
\end{center}
\caption{An INT/WFC V-band image centered on And~XVII. North is to the
  top and east to the left. The image spans $\sim 6.5\arcmin$ on a
  side.  The 5th magnitude foreground star, HD3346, lies $\sim 10\arcmin$ 
  to the North on the adjacent detector and scattered light from this 
  causes the background gradient seen in the Figure.}
\end{figure}

The new stellar system was discovered in data taken as part of a
northern extension to the INT survey, carried out during 23-29
September 2005.  Although the stellar concentration is just visible on
the V and $i$-band images (see Figure~1), the system is readily seen
on a map of stellar sources with magnitudes and colors appropriate for
metal-poor RGB stars at the distance of M31 (Figure~2). As we will
discuss, the properties of this system are consistent with it being a
dSph galaxy and we thus follow the naming convention originally
devised by \cite{vdb72} and refer to it as Andromeda~XVII
(And~XVII). Located at ($\alpha$,$\delta$)$_{2000}\approx
(00^h,37^m,07^s,44\arcdeg,19\arcmin,20\arcsec$), And~XVII lies at a
projected distance of 3.2$\arcdeg$ from the center of M31 making it,
in projection, one of the innermost known members of the satellite
system.

We note that although this is a relatively bright satellite (see 
Table~\ref{tab1}) located in quite a low density region around M31, the 
proximity of the 5th magnitude star HD344, and its associated scattered 
light, would have made detection on earlier photographic data quite 
difficult.  It is therefore unsurprising that discovery of such a system
required the advent of wide-field digital surveys.

\section{The Properties of And~XVII}

\input{tab1.tex}
The derived properties of And~XVII are summarised in Table~\ref{tab1}.
Figure~2 shows a contour map of a $0.5^\circ \times 0.5^\circ$ region
centred on the location of And~XVII for all stellar objects with magnitudes 
and colours consistent with metal-poor RGB stars at the distance of M31.
The overdensity corresponding to the dwarf is well-defined and clearly
separated from both foreground and M31 background components.
The contour map was constructed by binning the distribution
into 14$\arcsec$ pixels and then smoothing with a gaussian kernel of
FWHM$=70\arcsec$.  And~XVII is clearly extended along roughly an E-W direction 
with a measured average overall ellipticity of $\sim 0.2$.  In
the outer regions, the contours become significantly more elongated,
possibly indicative of tidal effects, but given the small number of
stars in these regions deeper observations are required to confirm this 
behaviour and examine their overall extent.

\begin{figure}[h!]
\begin{center}
\includegraphics[width=70mm,angle=-90]{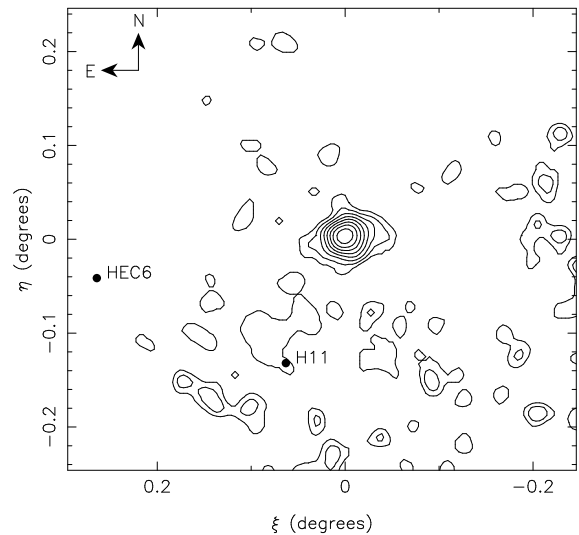}
\end{center}
\caption{A contour map of the distribution of stellar objects with
  magnitudes and colors consistent with metal-poor RGB stars at the
  distance of M31 with the positions of the nearby outer clusters 
  H11 and HEC6 indicated, The
  distribution has been binned into $14\arcsec$ pixels and then
  smoothed with a gaussian kernel of FWHM $70\arcsec$.  The first two
  contour levels are $\approx$2-sigma and 3-sigma and then increase
  at progressively larger increments to maintain visibility of the central 
  regions of the dwarf. The blank region to the top right is caused by the
  heavily saturated 5th magnitude star, HD344.}
\end{figure}

The tip of the RGB is a standard candle in old, low-metallicity
systems and thus can be used to estimate the line-of-sight distance to
the dwarf.  Although sparsely-populated, the luminosity function (LF)
of RGB stars in the vicinity of And~XVII shows clear evidence for a
discontinuity at I$_0\sim20.5\pm0.1$ (see Figure~3).  Assuming
$M_I(\rm{TRGB})=-4.04$ \citep{bella01,mc05} over the applicable
metallicity range, this leads to a distance modulus of
(m-M)$_0=24.5\pm0.1$ (D$=794\pm40$~kpc).  To within the errors,
And~XVII therefore lies at the same distance as M31.  Due to
background gradients induced by the nearby star HD344, the luminosity of
And~XVII proved impossible to derive reliably by direct integration.
We therefore opted to compare the upper RGB LF with that of And~V and
And~XI which we had also imaged on the INT/WFC using the same setup.
Making the plausible assumption of similar stellar populations for the
3 systems, direct comparison of the derived LFs yields an estimate of
the relative luminosity of And~XVII to be M$_V\sim -8.5$, i.e. of
comparable luminosity to And~IX but roughly 1 magnitude fainter than
And~V.

Figure~3 shows dereddened color-magnitude diagrams (CMDs) of stellar
sources lying within an ellipse of semi-major axis 2.5$\arcmin$,
ellipticity 0.2 and position angle 103$\arcdeg$ centered on And~XVII
and of stellar sources in a nearby comparison region of the same area.
A clear RGB sequence is apparent in the region coincident with
And~XVII but there is no evidence for any young main sequence
population. Globular cluster fiducials of metallicity [Fe/H]$=-0.71$
(47 Tuc), $-1.29$ (NGC~1851) and $-1.91$ (NGC~6397) are overlaid and
it can be seen that the system is metal poor, with a metallicity
best-matched by NGC~6397. 

\begin{figure}[h!]
\begin{center}
\includegraphics[width=50mm,angle=-90]{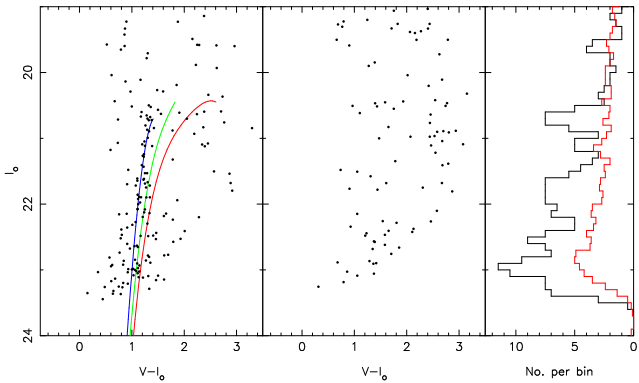}
\end{center}
\caption{Left and central panels: extinction-corrected CMDs of stars within 
  an elliptical region (a=$2.5\arcmin$,$e=0.2$ and PA$=103\arcdeg$) centered on
  And~XVII (left) and a nearby comparison region covering the same
  area (right). Overlaid on the And~XVII CMD are isochrones
  corresponding to 47~Tuc ([Fe/H]$=-0.7$), NGC~1851 ([Fe/H]$=-1.3$ and
  NGC~6397 ([Fe/H]$=-1.9$).  The rightmost panel shows the I-band LF of 
  the region centered on And~XVII compared to an offset region covering 5 times
  the area and renormalised appropriately.}
\end{figure}

The background-corrected radial profile of And~XVII was calculated in
elliptical annuli with PA and ellipticity held constant at the values
given in Table~\ref{tab1} and is shown in Figure~4.  The background level
of 0.9$\pm0.1$ stars per square arcmin was derived from the asymptotic level
attained in the outer parts.  The error from this is added in quadrature
to the Poisson count uncertainty to give the error bars shown in the Figure.

Overplotted are fits to a Plummer law and an exponential profile,
yielding scalelengths of 1.1$\arcmin$ (250 pc) and 0.65$\arcmin$ (150
pc) respectively.  These equate to half-light radii of 1.1$\arcmin$
(250 pc) in both cases.  The integral of the profile defined by the
Plummer model, coupled with our estimate of the absolute magnitude of
this system, defines the peak brightness in the Plummer model and
implies an extinction-corrected central surface brightness of 
$\mu_{\rm 0,V}\sim 26.1$ magnitudes per square arcsecond. Both Plummer and 
exponential fits provide a reasonably good match to the observed stellar 
profile to $\sim 3\arcmin$, beyond which the star counts may flatten (see
Figure~4). Although background uncertainties could be partly
responsible for this behaviour, such profile flattenings are often
seen at large radii in dSphs and could also result from a population
of extratidal stars \citep[e.g.][]{ih95, choi02}

\begin{figure}[h!]
\begin{center}
\includegraphics[width=65mm,angle=-90]{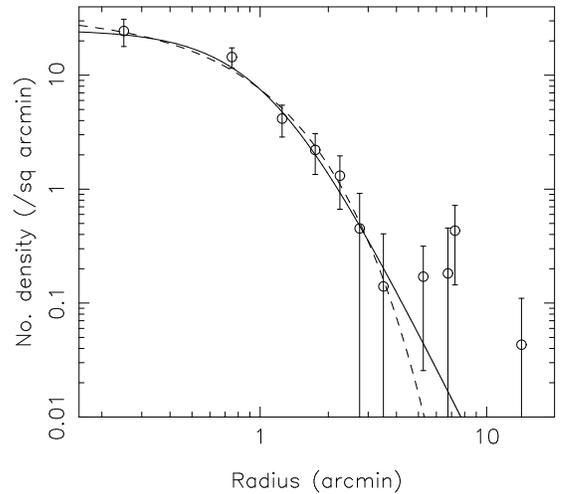}
\end{center}
\vspace{-3mm}
\caption{The background-corrected radial profile of And~XVII
  calculated from elliptical aperture photometry ($e=0.2$ and
  PA$=103\arcdeg$).  A background level of 0.9 stars per square armin
  is used. Overplotted are Plummer law (solid line) and exponential
  (dashed line) fits with scalelengths of $1.1\arcmin$ and
  $0.65\arcmin$ respectively.}
\end{figure}

\section{Discussion}

The overall properties we have derived for And~XVII are not unusual when
compared to those of other recently-discovered satellites of M31 and
the Milky Way.  Although low luminosity, And~XVII is $\sim 4-6$ times
more luminous than the ultra-faint M31 dSphs discovered by
\cite{martin06}. Furthermore, the metallicity of And~XVII is
consistent with that of similarly luminous M31 dSphs (e.g. And~IX,
And~X and And~XIV \citep{zucker04,zucker07, majewski07}).

There are, however, two properties of And~XVII which are of further
note.  First, quite unusually, there are three recently-discovered
globular clusters (GCs) which lie close to the main body of
And~XVII.  The projected distances to H11, HEC6 and HEC3 are
8.8$\arcmin$ (2~kpc), 16.1$\arcmin$ (3.7~kpc) and 25.7$\arcmin$
(5.9~kpc) respectively \citep{huxor08}.  While H11 is a classical
GC, HEC3 and HEC6 are so-called ``extended clusters''
which have luminosities typical of classical GCs but half-light
radii several times larger ($\gtrsim 30$~pc) \citep{huxor05,mackey06}.
This class of object has so far only been identified in M31, where 13
examples are currently known \citep{huxor08}, however the faintest
such objects are similar to the Milky Way's Palomar clusters. The fact
that two extended clusters lie within 6~kpc of And~XVII is therefore
extremely intriguing.  Interestingly, HEC6 lies close to
the projected major axis of the dwarf, which may be indicative of the
influence of tidal forces.

Figure 5 shows the distribution of dwarf satellites and GCs in the
outer halo of M31. Only the GCs identified by \cite{huxor08} and
\cite{martin06} are shown, however these represent essentially all the
confirmed clusters in the outer halo of M31 and have an average
surface density of $\approx$0.3 per square degree beyond a radius of 3
degrees. The group of 3 GCs within 30$\arcmin$ of And~XVII is
therefore quite unusual, with a probability of chance alignment of
$\approx$0.2\% (similar to the chance alignment of 2 of the GCs within
16$\arcmin$) and suggests that at least one of these objects may be
associated with the dwarf.  While several of the more luminous dwarf
elliptical companions of M31 possess their own GC systems
\citep[see][]{vdb00}, the only other M31 dSph with a GC projected
within 30$\arcmin$ is And~XIII. In this case, the GC lies $\sim
15\arcmin$ from the center of the dwarf but very much in the
foreground of M31 with a line-sight-distance of $631\pm58$~kpc
\citep{martin06}. No published distance estimate is yet available for
And~XIII though recently acquired Subaru imaging suggests it lies well
beyond M31 \citep{chapman08}.  Within the Milky Way dSph population,
only Fornax and Sagittarius possess their own GC systems, however
these are both considerably more luminous than And~XVII.  Radial
velocities and deeper CMDs are required to test the reality of the
association of And~XVII to the neighbouring star clusters.  If even
one of these is confirmed, it would imply a specific
frequency\footnote{S$_N=$N$_{GC} \times 10^{0.4(M_V+15)}$, where
  $N_{GC}$ is the number of GCs and M$_V$ is the host galaxy apparent
  magnitude \citep{hvdb81}.} of $\sim400$, making And~XVII very
unusual compared to other low luminosity galaxies, albeit with the
caveat that this may still be consistent with a low overall
probability of cluster formation in faint dwarf galaxies.

\begin{figure} 
\begin{center}
\includegraphics[width=90mm]{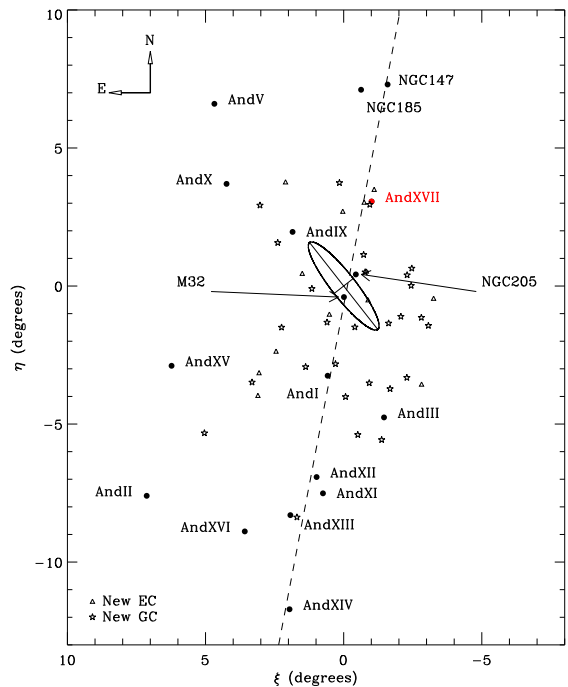}
\caption{A map of the distribution of currently-known dwarf satellites
and outer globular clusters of M31.  Compact clusters are represented
by asterixes and extended clusters by triangles. The inner ellipse has
a semi-major axis of 27~kpc and represents the bright disk of
M31. More than half of the known satellites of M31 lie within
$\lesssim 50\arcmin$ of a radial vector which passes through the
center of the galaxy and extends for $\gtrsim 19\arcdeg$ (dashed
line).}
\end{center}  
\end{figure}

Another curious aspect of And~XVII is its projected position on the
sky. Patterns in the spatial distribution of satellites around host
galaxies have long been a subject of much interest
\citep[e.g.][]{holm69} and previous analyses of the M31 system have
been reported by \cite{kara96}, \cite{mc06} and \cite{koch06}. Given
the rapid improvement in our census of M31 satellites in the last few
years, even these recent studies are based on fairly incomplete
samples.  Considering projected positions on the sky alone,
\cite{majewski07} have recently commented on the fact that many of the
known satellites of M31 lie along a vector which connects And~XIV to
NGC~185 and NGC~147. Remarkably, And~XVII is also consistent with this
alignment.  The result from a linear least-squares fit to the
projected positions of NGC~147, NGC~185, And~XVII, NGC~205, M32,
And~I, And~XII, And~XI, And~XIII and And~XIV is shown in Figure 4.  Of
the 10 satellites (out of the 19 currently-known), none deviate by
more $50\arcmin$ ($<12$~kpc) from this vector over its $19\arcdeg$
($260$~kpc) extent.  However, more accurate distances and further
radial velocities are  needed to test if this alignment is
a physical association rather than a chance projection effect,
compounded by inhomogeneous survey depths.

Full analysis of the satellite distribution around M31 requires a
comprehensive search in the remaining quadrants.  We are currently
using MegaCam on CFHT to survey an additional quadrant to a radius of
$\sim 150$~kpc. Further, the Pan-Starrs 3$\pi$ survey is due to get
underway in late 2008 and although only reaching depths comparable to
the INT/WFC survey, will do so over a much larger area.  The prospects
are therefore extremely promising for
developing a more complete picture of the satellite and globular
cluster system of M31.

\acknowledgments

We would like to thank the referee for their careful reading of the manuscript
and helpful comments. AMNF and AH are supported by a Marie Curie 
Excellence Grant from the European Commission under contract 
MCEXT-CT-2005-025869. NRT acknowledges financial support via an STFC Senior 
Research Fellowship.


\end{document}

%% file: tab1.tex
\begin{deluxetable}{lc}
\tablecaption{Properties of And~XVII \label{tab1}}
\tablewidth{0pt} \tablehead{ \colhead{Parameter\tablenotemark{a}} &
{~~~ } } \startdata Coordinates (J2000) & 00:37:07.0\, 44:19:20\\
Coordinates (Galactic) & $\ell = 120.23^\circ, b = -18.47^\circ$ \\
 Position Angle & $\approx 103^{\circ}$\\
 Ellipticity & $\approx 0.2$\\
 $r_h$ (Plummer) & $1\farcm1\pm0.1$\\
 A$_{\rm V}$ & $0\fm246$ \\
 $\mu_{\rm 0,V}$ (Plummer) & $26\fm1$\\
 $V_{\rm tot}$ & $16\fm0$\\
(m$-$M)$_0$ & $24\fm5$\\
 M$_{\rm tot,V}$ & $-8\fm5$
\enddata
\tablenotetext{a}{Surface brightnesses and integrated magnitudes 
are accurate to $\sim \pm 0.5$~mag and are corrected for the 
mean Galactic foreground reddening, A$_{\rm V}$, shown.}
\label{tab:struct}
\end{deluxetable}

%% file: astroph.bbl
\begin{thebibliography}{}
\bibitem[Armandroff et al.(1998)]{armand98} Armandroff, T.~E., Davies, J.~E., \& Jacoby, G.~H.\ 1998, \aj, 116, 2287 
\bibitem[Armandroff et al.(1999)]{armand99} Armandroff, T.~E., Jacoby, G.~H., \& Davies, J.~E.\ 1999, \aj, 118, 1220 
\bibitem[Bellazzini et al.(2001)]{bella01} Bellazzini, M., Ferraro, F.~R., \& Pancino, E.\ 2001, \apj, 556, 635 
\bibitem[Belokurov et al.(2007)]{belo07} Belokurov, V., et al.\ 2007, \apj, 654, 897 
\bibitem[Belokurov et al.(2006)]{belo06} Belokurov, V., et 
al.\ 2006, \apjl, 647, L111
\bibitem[Chapman et al.(2008)]{chapman08} Chapman, S., et al.\ 2008, in preparation.
\bibitem[Choi et al.(2002)]{choi02} Choi, P.~I., Guhathakurta, P., \& Johnston, K.~V.\ 2002, \aj, 124, 310 
\bibitem[Ferguson et al.(2000)]{ferg00} Ferguson, A.~M.~N., Gallagher, J.~S., \& Wyse, R.~F.~G.\ 2000, \aj, 120, 821 
\bibitem[Ferguson et al.(2002)]{ferg02} Ferguson, A. M. N., Irwin, M. J., Ibata, R. A., Lewis, G. F., \& Tanvir, N. R. 2002, AJ, 124, 1452
\bibitem[Harris \& van den Bergh(1981)]{hvdb81} Harris, W.~E., \& van den Bergh, S.\ 1981, \aj, 86, 1627 
\bibitem[Holmberg(1969)]{holm69} Holmberg, E. 1969, Ark. Astron., 5, 305
\bibitem[Huxor et al.(2005)]{huxor05} Huxor, A. P., Tanvir, N. R., Irwin, M. J., Ibata, R., Collett, J. L., Ferguson, A. M. N., Bridges, T., \& Lewis, G. F. 2005, MNRAS, 360, 1007
\bibitem[Huxor et al.(2008)]{huxor08} Huxor, A. P., Tanvir, N. R., Ferguson, A. M. N., Irwin, M. J., Ibata, R., Bridges, T., \& Lewis, G. F. 2008, MNRAS, in press.
\bibitem[Ibata et al.(2001)]{ibata01} Ibata, R., Irwin, M., Lewis, G., Ferguson, A.~M.~N., \& Tanvir, N.\ 2001, \nat, 412, 49 
\bibitem[Ibata et al.(2007)]{ibata07} Ibata, R., Martin, N.~F., Irwin, M., Chapman, S., Ferguson, A.~M.~N., Lewis, G.~F., \& McConnachie, A.~W.\ 2007, \apj, 671, 1591 
\bibitem[Irwin \& Hatzidimitriou(1995)]{ih95} Irwin, M., \& Hatzidimitriou, D.\ 1995, \mnras, 277, 1354 
\bibitem[Irwin \& Lewis(2001)]{irwin01} Irwin, M., \& Lewis, J.\ 2001, New Astronomy Review, 45, 105 
\bibitem[Irwin et al.(2005)]{irwin05} Irwin, M.~J., Ferguson, A.~M.~N., Ibata, R.~A., Lewis, G.~F., \& Tanvir, N.~R.\ 2005, \apjl, 628, L105 
\bibitem[Irwin et al.(2007)]{irwin07} Irwin, M.~J., et al.\ 2007, \apjl, 656, L13 
\bibitem[Karachentsev(1996)]{kara96} Karachentsev, I.\ 1996, \aap, 305, 33 
\bibitem[Karachentsev \& Karachentseva(1999)]{kara99} Karachentsev, I.~D., \& Karachentseva, V.~E.\ 1999, \aap, 341, 355 
\bibitem[Koch \& Grebel(2006)]{koch06} Koch, A., \& Grebel, E.~K.\ 2006, \aj, 131, 1405 
\bibitem[Komiyama et al.(2007)]{komiyama07} Komiyama, Y., et al.\ 2007, \aj, 134, 835 
\bibitem[Koposov et al.(2007)]{koposov07} Koposov, S., et al.\ 2007, \apj, 669, 337 
\bibitem[Landolt(1992)]{land92} Landolt, A.~U.\ 1992, \aj, 104, 340 
\bibitem[Mackey et al.(2006)]{mackey06} Mackey, A.~D., et al.\ 2006, \apjl, 653, L105 
\bibitem[Majewski et al.(2007)]{majewski07} Majewski, S.~R., et al.\ 2007, \apjl, 670, L9 
\bibitem[Martin et al.(2006)]{martin06} Martin, N. F., Ibata, R. A., Irwin, M. J., Chapman, S., Lewis, G. F., Ferguson, A. M. N., Tanvir, N., \& McConnachie, A. W. 2006, MNRAS, 371, 1983
\bibitem[McConnachie et al.(2004)]{mc04} McConnachie, A.~W., Irwin, M.~J., Ferguson, A.~M.~N., Ibata, R.~A., Lewis, G.~F., \& Tanvir, N.\ 2004, \mnras, 350, 243 
\bibitem[McConnachie et al.(2005)]{mc05} McConnachie, A.~W., Irwin, M.~J., Ferguson, A.~M.~N., Ibata, R.~A., Lewis, G.~F., \& Tanvir, N.\ 2005, \mnras, 356, 979 
\bibitem[McConnachie \& Irwin(2006)]{mc06} McConnachie, A.~W., \& Irwin, M.~J.\ 2006, \mnras, 365, 902 
\bibitem[McConnachie et al.(2007)]{mc07} McConnachie, A.~W., Arimoto, N., \& Irwin, M.\ 2007, \mnras, 379, 379 
\bibitem[Morrison et al.(2003)]{morrison03} Morrison, H.~L., Harding, P., Hurley-Keller, D., \& Jacoby, G.\ 2003, \apjl, 596, L183 
\bibitem[Sakamoto \& Hasegawa(2006)]{saka06} Sakamoto, T., \& Hasegawa, T.\ 2006, \apjl, 653, L29 
\bibitem[Schlegel et al.(1998)]{schlegel98} Schlegel, D. J., Finkbeiner, D. P., \& Davis, M. 1998, ApJ, 500, 525
\bibitem[Simon \& Geha(2007)]{geha07} Simon, J.~D., \& Geha, M.\ 2007, \apj, 670, 313
\bibitem[van den Bergh(1972)]{vdb72} van den Bergh, S.\ 1972, \apjl, 171, L31
\bibitem[van den Bergh(2000)]{vdb00} van den Bergh, S.\ 2000, The Galaxies of the Local Group, by Sidney Van den Bergh.~Published by Cambridge, UK: Cambridge University Press, 2000 Cambridge Astrophysics Series, Vol 35   
\bibitem[Walsh et al.(2007)]{walsh07} Walsh, S.~M., Jerjen, H., \& Willman, B.\ 2007, \apjl, 662, L83 
\bibitem[Willman et al.(2005a)]{willman05a} Willman, B., et al.\ 2005a, \apjl, 626, L85 
\bibitem[Willman et al.(2005b)]{willman05b} Willman, B., et al.\ 2005b, \aj, 129, 2692 
\bibitem[Zucker et al.(2004)]{zucker04} Zucker, D.~B., et al.\ 2004, \apjl, 612, L121 
\bibitem[Zucker et al.(2006a)]{zucker06a} Zucker, D.~B., et al.\ 2006a, \apjl, 650, L41 
\bibitem[Zucker et al.(2006b)]{zucker06b} Zucker, D.~B., et al.\ 2006b, \apjl, 643, L103
\bibitem[Zucker et al.(2007)]{zucker07} Zucker, D.~B., et al.\ 2007, \apjl, 659, L21 
\end{thebibliography}
